# Diverse regimes of mode intensity correlation in nanofiber random lasers through nanoparticle doping

Martina Montinaro[1], Vincenzo Resta[1], Andrea Camposeo[2,**], Maria Moffa[2], Giovanni Morello[1],

Luana Persano[2], Karolis Kazlauskas[3], Saulius Jursenas[3], Ausra Tomkeviciene[4],

Juozas V. Grazulevicius[4], Dario Pisignano[2,5,*]

[1] Dipartimento di Matematica e Fisica "Ennio De Giorgi", Università del Salento, via Arnesano, I-73100 Lecce, Italy

[2] NEST, Istituto Nanoscienze-CNR, Piazza San Silvestro 12, I-56127 Pisa, Italy

[3] Institute of Applied Research, Vilnius University, Saulėtekio 3, LT-10257 Vilnius, Lithuania

[4] Department of Polymer Chemistry and Technology, Kaunas University of Technology, Radvilenu Plentas 19, LT-50254 Kaunas, Lithuania

[5] Dipartimento di Fisica, Università di Pisa, Largo B. Pontecorvo 3, I-56127 Pisa, Italy








ABSTRACT. Random lasers are based on disordered materials with optical gain. These devices can exhibit either intensity or resonant feedback, relying on diffusive or interference behaviour of light, respectively, which leads to either coupling or independent operation of lasing modes. We study for the first time these regimes in complex, solid-state nanostructured materials. The number of lasing modes and their intensity correlation features are found to be tailorable in random lasers made of light-emitting, electrospun polymer fibers upon nanoparticle doping. By material engineering, directional waveguiding along the length of fibers is found to be relevant to enhance mode correlation in both intensity feedback and resonant feedback random lasing. The here reported findings can be used to establish new design rules for tuning the emission of nano-lasers and correlation properties by means of the compositional and morphological properties of complex nanostructured materials.






Disordered materials with optical gain are the building blocks of random lasers, whose operation is based on the scattering properties of light.[1] These devices can be useful for a wide range of applications, which include diagnostic[2] and speckle-free[3] imaging, spectroscopic tools for monitoring biomaterials and structural deformations,[4] new laser projector schemes and optical tomography.[5] A number of experimental[6-9] and theoretical[10-13] approaches have been developed to rationalize the behaviour and to tailor the performances of random lasers. In this framework, two main classes of devices are distinguished and experimentally observed. In so-called intensity (or incoherent) feedback random lasing (IFRL), the propagation of light can be described as a diffusional process with amplification, neglecting interference effects.[10,14] This mechanism leads to smooth and narrow emission peaks, with full width at half maximum (FWHM) of a few nm. Instead, in resonant feedback random lasing (RFRL), interference upon multiple scattering plays a major role, which might lead to some degree of spatial localization of the light modes, and to very narrow (FWHM down to sub-nm) lasing peaks, as found since 1999[6] in numerous systems at the solid state. The RFRL to IFRL transition has been recently investigated by reducing the directionality of pumping photons.[9,15] In this way, this transition could be associated to the coupling of modes which determines a condensation-like process, namely the onset of collective oscillations from μm-sized titania clusters in a rhodamine solution. Extending the study of this transition to solid-state devices would be highly important to understand how the emission features and mode interactions of random lasers can be tailored. Indeed, regimes with different feedback or inter-mode coupling should be obtained not only by shaping the excitation beam as in previous reports, but also by varying the degree of disorder, through the composition and morphology of materials showing gain properties.[16] These include a large variety of systems such as conjugated polymers,[8] dyes in plastic matrices, sub-wavelength particles with plasmon-enhanced scattering,[17] cellulose fibers doped with Au nanoparticles,[18] and light-emitting nanocomposites,[19] usable





in displays, organic lasers, and in nanophotonics.

Here we show how spectral characteristics and mode intensity correlation features can be tailored in random lasers made of electrospun polymer fibers with optical gain upon controlling local disorder by doping with TiO$_2$ nanoparticles. Fibers serve as multifunctional component in the devices, providing stimulated emission, waveguiding along their longitudinal axis (multidirectional in disordered mats), and moderate scattering due to their refractive index (~1.6) and cylindrical body.[20] These properties induce IFRL with highly interacting resonances, resulting in a strongly correlated random lasing emission. Clustered TiO$_2$ particles provide additional elastic scattering and feedback in the complex material, and are found to drive the system to RFRL with discrete and more weakly interacting lasing modes. The onset of the RFRL-to-IFRL transition is also probed reducing guided radiation, through directionality selection for collected modes. These results are the first experimental evidence of the role of waveguiding in enhancing mode coupling in random lasers.

**Results and Discussion**

We fabricate random lasers made of disordered mats of electrospun polystyrene (PS) fibers (Fig. 1a) doped with the recently proposed 2,7-bis(9,9-diethylfluoren-2-yl)-9-(2-ethylhexyl)carbazole (Fl-Cz-Fl) as gain material[21] and with a further degree of disorder provided by optionally embedded TiO$_2$ particles (Fig. 1b). PS is chosen as thermoplastic polymer with refractive index about 1.6 and good optical transparency in the visible and near infrared, leading to waveguides with low propagation losses and improved mode confinement compared to other transparent polymers.[22] In addition, the elastic modulus (of the order of 1 GPa) and the thermal properties (glass-transition temperature, $T_g$~100 °C) of electrospun PS fibers[23] make plastic lasers based on them stable while operating at room temperature. Fl-Cz-Fl is a blue-emitting compound featuring fluorene and carbazole moieties arranged into 2,7-





substitution pattern, and shows high emission quantum yield (~0.8) and large radiative decay rate (~$10^9$ $s^{-1}$).[21] Additionally, the twisted molecular structure and bulky peripheral moieties of Fl-Cz-Fl lead to high $T_g$ (68 °C) ensuring formation of stable amorphous phase. The electrospinning technique is very effective in producing almost one-dimensional polymer structures, including light-emitting nanofibers,[24] given that the used electrified solution shows a sufficient amount of macromolecule entanglements.[25,26] Hybrid fibers can be straightforwardly electrospun, incorporating nanoparticles in organic filaments to tailor optical properties through the embedded components. Also, the method is highly convenient to realize random lasers made of light-emitting fibers with optical gain and assembled in slabs. Indeed, the critical thickness for lasing from slabs is given by $t_{CR} = \pi(l_G\, l_T/3)^{1/2}$, where $l_G$ and $l_T$ indicate the gain length and the transport mean free path of the diffused photons, respectively.[10] For lasing dyes in plastic matrices, $l_G$ might be of the order of a few hundreds of μm, whereas $l_T$ in our membranes of electrospun fibers is estimated in the range 2-10 μm by coherent backscattering (CBS) experiments. This leads to $t_{CR}$ values of about 100 μm, which are well within the range of thickness which is easily obtained by electrospinning. The here realized organic fibers show an average diameter of about 4 μm (Fig. S1a in the Supporting Information), with micron-size roughness and roughly cylindrical body as shown in Fig. 1a. When inserted in the electrospun solution, $TiO_2$ particles (~15-23 nm, Fig. S1b) can conglomerate within fibers, forming clusters with size up to a few μm (Fig. 1b), thus introducing additional elastic scattering for transported light. Exemplary photographs of the electrospun mats under white illumination conditions and under UV excitation are shown in Fig. 1c and 1d, respectively. The different sample morphologies and compositions can so lead to varied emission characteristics, as shown in Fig. 1e,f. Upon excitation with about 10 ns pulses at 355 nm and with a stripe excitation geometry (see Methods), lasing spectra from samples without $TiO_2$ particles exhibit a single peak at about 417 nm with FWHM of about 6 nm. The spectra are mostly





smooth and only slightly featured at their top, and are ascribed to IFRL. Instead, emission spectra from fibers doped with $TiO_2$ particles are spiky, with several narrow (FWHM $\cong$ 0.3 nm) peaks whose wavelength is quite stable upon changing excitation fluence. These features are clearly to be ascribed to RFRL. Well-defined excitation thresholds are found for lasing emission, at about 15 mJ cm$^{-2}$ and at about 20 mJ cm$^{-2}$ for purely organic and for $TiO_2$-doped fibers, respectively, while the background spontaneous emission, i.e. photoluminescence (PL) signal grows weakly and largely linearly upon increasing the excitation fluence. One might also expect a significant contribution to mode excitation coming from indirect pumping by amplified spontaneous emission (ASE) from the surrounding regions, which is waveguided in the organic slab constituted by the polymer non-woven. The persistence of ASE components directionally guided along networks of randomly-oriented nanofibers has been shown in previous work.[27] The increase of the lasing threshold upon embedment of the $TiO_2$ particles is therefore related to the correspondingly decreased characteristic scattering mean free path, experienced by the directional pumping from ASE along the excitation stripe.[28] Various methods would allow the excitation threshold of fiber-based random lasers to be decreased, including the exploitation of plasmon-enhanced light scattering by incorporated Au nanoparticles.[18] Overall, the addition of $TiO_2$ nanoparticles and the increased local degree of disorder clearly drive the fiber-based random laser from a multimode IFRL regime to the RFRL regime, featuring few discrete lasing modes. In this respect, this transition from a smoother to a more spiky emission is here demonstrated to be associated in a clear way with the compositional and morphological complexity of the disordered medium, offering a direct method to tailor spectral characteristics by material engineering.

Two different scattering components are present in $TiO_2$-doped fiber mats, which are the polymer filaments with highly elongated shape[20] and with waveguiding capability along their axis,[29] and the inorganic particles, respectively. To better rationalize the contribution of the different scatterers,





we consider the correlation of intensities of different modes in the emission spectra from samples with and without TiO$_2$ nanoparticles. The Pearson's correlation coefficient, $C$, for the intensity fluctuations at a couple of wavelengths ($\lambda_j$, $\lambda_k$) is:

$$C(\lambda_j, \lambda_k) = \frac{\sum_{i=1}^{N}[I_i(\lambda_j)-\bar{I}(\lambda_j)][I_i(\lambda_k)-\bar{I}(\lambda_k)]}{\sqrt{\sum_{i=1}^{N}[I_i(\lambda_j)-\bar{I}(\lambda_j)]^2}\sqrt{\sum_{i=1}^{N}[I_i(\lambda_k)-\bar{I}(\lambda_k)]^2}}, \quad (1)$$

where $\bar{I}$ indicates the average over the ensemble given by $N$ analyzed shots (excitation pulses). The $C$ coefficient is zero for completely uncorrelated peak intensities, and reaches a unity maximum value when peaks are fully correlated. The inter-mode spectral correlation is here studied for $j,k$=1,..,5, namely for five relevant wavelengths highlighted by arrows in the insets of Fig. 1e and 1f. The emission from our interconnected network of fibers under directional excitation generally exhibits correlation coefficients (0.76-0.97), which are significantly larger than those reported for uncorrelated, RFRL modes in solutions with scattering nanoparticles.[9] The role of waveguiding filaments in promoting mode intensity correlation is also supported by the higher average correlation found for Fl-Cz-Fl/PS fibers (0.94±0.02) compared to mats with scattering particles (0.89±0.05). These results are shown in the correlation matrix plots in Fig. 2a and in Fig. S2a, respectively. Here the normalized histograms showing the frequencies of the measured intensities at fixed wavelengths (from shot to shot), are plotted along the matrix diagonal, whereas the off-diagonal panels in Fig. 2a display intensity data interdependencies of selected couples of modes, evidencing high correlation in each pair of intensities, with $C$ values close to unity. Instead, scattering particles lead to a de-correlation of the overall emission spectra (Fig. S2), and to an increase of the spread of the intensity data around the diagonal line (Fig. S3 and S4). This is also reflected in the relatively higher standard deviations of the correlations found for samples with nanoparticles, which indicates that $C$ significantly depends on the pairs of wavelengths chosen, namely on the specific modes involved. In Fig. 2b, we show the shot-to-





shot evolution of emission spectra for Fl-Cz-Fl/PS fibers. Established modes have quite stable intensity at different excitation shots, whereas for TiO$_2$-doped fibers one finds larger fluctuations over time, suggesting a stochastic dependence in the onset of resonances (Fig. S2b).

To better highlight how $C$ depends on wavelengths, we plot two-dimensional correlation maps for the spectra of Fl-Cz-Fl/PS fibers without and with TiO$_2$ in Fig. 3a and 3b, respectively. The highly correlated spectral region of the fiber random laser is clearly appreciable around the autocorrelation diagonal (top-left to bottom-right map angles in Fig. 3a). Upon embedding TiO$_2$ nanoparticles into the mat of light-emitting fibers, the map becomes textured in a very different way, featuring a grid pattern superimposed to the autocorrelation diagonal (Fig. 3b). This highlights lower inter-correlation of the main peak wavelengths, suggestive of relatively poor interaction established between spectrally adjacent modes. These findings are in agreement with recent results reported for the random lasing fluctuations in gelatinous gain media with colloidal scatterers.[30] Indeed, points defined by crossing vertical and horizontal lines are highly correlated only if belonging to the autocorrelation diagonal (red dots in the map in Fig. 3b), whereas adding particles in the laser materials leads to the emergence of discrete pairs of lasing peaks with correlations values < 0.5 (cyan, off-diagonal dots in Fig. 3b), namely with modest intensity correlation.[31] Squared regions with vertexes in such dots exhibit high correlations within their area and a lower one on their sides, as exemplified by the rectangle displayed in the map.

The correlation and spectral features can be also conveniently analyzed by covariance mapping.[32] To this aim we plot the covariance ($\sum_{i=1}^{N}[I_i(\lambda_j) - \bar{I}(\lambda_j)][I_i(\lambda_k) - \bar{I}(\lambda_k)])/N$) for emission intensities at each pair of wavelengths (Fig. 3c,d), which directly indicates how the two variables $I(\lambda_j)$ and $I(\lambda_k)$ change together from shot to shot. For Fl-Cz-Fl/PS fibers, the resulting map has a symmetric shape (Fig. 3c), where concentric ellipses of different colours correspond to different levels of emission intensity (i.e. the decrease of intensity at the tails of the lasing peak occurs in a similar way for any





shot). Instead, the grid pattern found for the *C*-map is even more clear in the covariance map for lasing fibers doped with TiO$_2$ (Fig. 3d).

For random lasers made of gain solutions with nanoparticle scatterers, excited with spots of different shapes, a mode-locking transition has been associated with the change from resonant to incoherent feedback regimes, and with the directionality of excitation and the spatial overlap of modes.[9] Spatially distant or weakly overlapping modes might interact only partly, which would results in low correlation. Here, with identical excitation geometry Fl-Cz-Fl/PS fibers exhibit either IFRL or RFRL spectral characteristics, depending on TiO$_2$ doping, while the intensity correlation is high on average for both the investigated samples, evidencing mode interaction taking a role even in the RFRL regime. This peculiar behaviour is due to various interplaying mechanisms which affect photon transport in the complex material composed by nanofibers. On one side, electrospun light-emitting polymer filaments might show, in addition to light-scattering whose form factor depends on the lasing wavelength and on the fiber cross-section,[20] significant waveguiding properties. In our set-up the excitation stripe is large enough to pump a large number of fibers (Fig. 4a), thus triggering different waveguided modes along various directions. The loss coefficient, $\gamma$, for self-emitted light propagating along the length of individual electrospun fibers are down to less than 100 cm$^{-1}$, corresponding to effective photon transport over many tens of $\mu$m,[29] which could favour coupling of otherwise poorly overlapping modes thus leading to correlated lasing modes. Fig. 4b shows an exemplary fluorescence microscopy image showing an excited Fl-Cz-Fl/PS fiber with propagating light exiting the two tips of the polymer filament, which are distant about 120 $\mu$m. On the contrary, TiO$_2$ particles disfavor photon transport along the organic material, possibly developing resonant cavities in the slab as a route to modal selection and consequently to very narrow lasing peaks. The intensity correlation of optical modes is correspondingly weakened. Through the formation of random ring resonators, a similar





behaviour has been found in dye-doped fibers with deterministic, comb-like spectral emission.[33] Unlike previous works, however, our system is not composed by well-established cavities generated during fiber deposition, featuring instead a complex network of organic or hybrid filaments with significant diffusive behaviour for transported light. Here the waveguiding effect is quantified by measuring the emission intensity as a function of the varying distance, $d$, of the excitation region from the emitting edge of aligned Fl-Cz-Fl/PS fibers as schematized in the bottom inset of Fig. 4c. The decrease of the output intensity upon increasing $d$ (top inset of Fig. 4c) is then fitted by a law, $I_{PL} = I_0\,e^{-\gamma d}$, where $I_0$ is a constant, providing a value of the loss coefficient as low as 7.2 cm$^{-1}$ (Fig. 4c). This value is comparable to propagation loss coefficients of active polymer slab waveguides,[34,35] that is indicative of effective light transport along the disordered network of fibers. Waveguiding over millimeter distances is likely favoured by both the optical confinement properties of individual fibers and by inter-fiber coupling. These properties of electrospun fibrous materials makes them highly interesting for random lasing applications, as well as for investigating coupling regimes promoted by enhanced light transport.

We design specific experiments to test further how waveguiding properties of the electrospun fibers affect inter-mode correlation. To this aim, photon transport along the fiber length is intentionally suppressed by embedding the mats (without TiO$_2$ doping) in a resin matrix, thus reducing the refractive index contrast ($\Delta n$) between the internal and the external regions of the cylindrical, dielectric waveguides. Leading to higher optical leakages from fibers, this is found to induce a relevant change in the covariance for emission intensities from different wavelengths. The resulting map is shown in Fig. 5a. Correspondingly, we find $C$ values close to unity which are signature of enhanced IFRL character. Furthermore, we calculate the covariance map for light emitted by random Fl-Cz-Fl/PS fibers (without TiO$_2$ particles and without resin), upon positioning the mat at an angle of 45° (± 2°) with respect to the collecting optical fiber (Fig. S5a). This method allows for better decoupling waveguided radiation,





propagating in fibers, and detected light, the latter thus being relatively enriched in the scattered

component. Again, the resulting map starts to resembling a grid pattern, namely it highlights relatively

de-correlated emission peaks similarly to RFRL devices incorporating scattering particles (Fig. 5b).

This is also supported by a decrease of the correlation coefficient to 0.82±0.09. Similar considerations,

showing a transition towards spikiness, hold for emission spectra (Fig. S5b).

Previous studies have shown that in dye solutions with titania particles, spiky lasing spectra are

largely restricted to weakly scattering systems, while multiple light scattering events or lower gain

more likely lead to smooth lasing spectra as in IFRL.[36] This is interpreted in terms of the diffusive and

omnidirectional character of random lasers with relatively stronger scattering, where many modes share

the same gain medium. As we evidence by changing the photon collection direction (Fig. 5b), in

random lasers with fibers relevant elements, including waveguiding, are present instead which

introduce directionality namely angular mode selectivity in the emission. On the other hand,

multidirectional waveguides constituted by the electrospun fibers are likely to provide the material with

enhanced modal correlation. Indeed, at variance with early works exploiting titanium dioxide powders

in solutions,[9,37] here highly correlated spectral modes are found even by a stripe excitation geometry.

Since such geometry is likely to lead to a lower intrinsic interaction of distant modes, locking different

emissive regions through the network of nanofibers might be critically important to establish

correlation in different and narrow peaks observed in this work. In this way, also RFRL peaks, in

principle originating from spatially decoupled positions, and not only IFRL regions, can undergo

collective oscillations. In this respect, the decrease of optical losses provided by light-emitting

nanofibers[29] can be therefore relevant to prevent the inhibition of the coupling dynamics[15] and to favour

a mode correlation which is significant also in RFRL regime. In addition, increasing the scattering

strength by adding nanoparticles may lead to a double effect including directionality suppression





(which would strengthen the IFRL character of the emission) as well as development of extra resonant cavities. These would interact through the network of waveguiding polymer filaments, leading the device to RFRL with significant though relatively weaker intensity correlation. Our results suggest that this latter mechanism, not present in experiments in solution,[9,37] might prevail in solid-state nanomaterials.

**Conclusions**

Overall, these findings show how tunable, multicomponent disordered morphologies might allow the intensity correlation of light modes to be tailored, activated or suppressed in hybrid nano- and microstructures. Controlling the transition from IFRL to RFRL character in random lasing materials represents an important step towards engineering light confinement and emission in these systems. The occurrence of a regime of mode-locked lasing with intensity-correlated modes was envisaged at low degrees of disorder, whereas increasing disorder should drive the lasing system towards modes coherent in phase but not in intensity.[16] Our work is a first experimental study focusing on mode-intensity correlation in structurally disordered, fibrous materials. In this framework electrospun fibers with controlled nanoparticle doping are a versatile platform, which allows morphology and composition of hybrid lasers to be varied in a controlled way, materials with effective waveguiding networks to be realized, and resonances to be tailored, which can establish new design rules for amorphous photonics.





**Methods**

*Light-emitting nanofibers.* Polymer fibers are electrospun from a solution of 300 mg/mL of PS (Sigma-Aldrich, molecular weight = 192,000 Da) in chloroform (Sigma-Aldrich), dissolving Fl-Cz-Fl at 5% (w/w) with respect to PS. The synthesis procedure of Fl-Cz-Fl has been reported elsewhere.[21] In parallel, another solution is prepared adding $TiO_2$ nanoparticles (Titanium(IV) oxide nanopowder, Sigma-Aldrich, at a concentration of 1% (w/w) with respect to PS. The Norland Optical Adhesive 68 (NOA68) is used to optionally embed electrospun fibers. Solutions are stirred for 12 h, loaded into a 1 mL syringe, delivered at a constant flow rate of 1 mL/h through a metal needle (21 gauge) by a peristaltic pump (33 Dual Syringe Pump, Harvard Apparatus Inc., Holliston, MA), and electrospun at ambient conditions. The needle is connected to a high-voltage power supply (EL60R0.6–22, Glassman High Voltage, High Bridge, NJ), applying a bias of 13 kV. Fibers are collected on a grounded metallic plate at a distance of 10 cm from the needle. Arrays of uniaxially aligned Fl-Cz-Fl/PS fibers are also fabricated by using a rotating collector (4000 rpm). For encapsulation, the fibrous mats are embedded NOA68, which is then cured by exposure to UV light (3 mW/cm$^2$) for 3 min. A sample consisting of a NOA68 layer sandwiched between two glass substrates and without fibers is prepared to measure the resin transmission (Fig. S5b).

*Morphological and optical characterization.* The morphology of electrospun fibers is investigated by scanning transmission electron microscopy (Nova NanoSEM 450, FEI) with an acceleration voltage of 30 kV. Fluorescence micrographs are collected by an inverted microscope (Nikon) working in epi-layer configuration. The fibers are excited by a mercury lamp, while the emission is collected by the same objective and imaged by a charge coupled device (CCD) camera. Confocal fluorescence micrographs are obtained by exciting the sample with a diode laser ($\lambda = 405$ nm), focused through an objective lens (20×, numerical aperture = 0.5). The emission from the fibers is





then collected by the same objective and analyzed by a multianode photomultiplier. A spectrophotometer (Perkin Elmer, Lambda 950) is used for transmittance measurements. The light mean free path of nanofibers mats is measured through CBS by a He:Ne laser beam with wavelength at 633 nm. The backscattered light intensity angular profile is measured by a CCD Camera (iDUS, Andor).

*Random lasers*. Random lasing is achieved by pumping samples with the third harmonic ($\lambda$ = 355 nm) of a Nd:YAG laser beam (repetition rate = 10 Hz, pulse duration $\cong$ 10 ns), focused in a rectangular stripe (55 $\mu$m×1.9 mm). The emission is then collected from the sample edge, coupled to an optical fiber and analyzed by a monochromator (iHR320, Jobin Yvon) equipped with a CCD detector (Symphony, Jobin Yvon). Random lasing thresholds are measured by systematically varying the excitation fluence, and series of single-shot spectra are acquired with an integration time of 50 ms. By keeping fixed the pump fluence, optical losses are measured through the emitted light intensity by moving the excitation stripe away from the sample emitting edge.





AUTHOR INFORMATION

**Corresponding Authors**

* dario.pisignano@unipi.it

** andrea.camposeo@nano.cnr.it

ACKNOWLEDGMENT

The research leading to these results has received funding from the European Research Council under the European Union's Seventh Framework Programme (FP/2007-2013)/ERC Grant Agreement n. 306357 (ERC Starting Grant "NANO-JETS"). The Apulia Networks of Public Research Laboratories WAFITECH (09) and MITT (13), and the Fund for Development and Cohesion 2007-2013 APQ Apulia Region Research "Regional program for smart specialization and social and environmental sustainaibility" (YP9DNY7) are also acknowledged. K. K. and S. J. acknowledge funding by a grant (No. LJB-3/2015) from the Research Council of Lithuania. V. Fasano is acknowledged for CBS measurements. A. Fornieri is acknowledged for scanning electron microscopy on nanoparticles.

## Figures and Captions

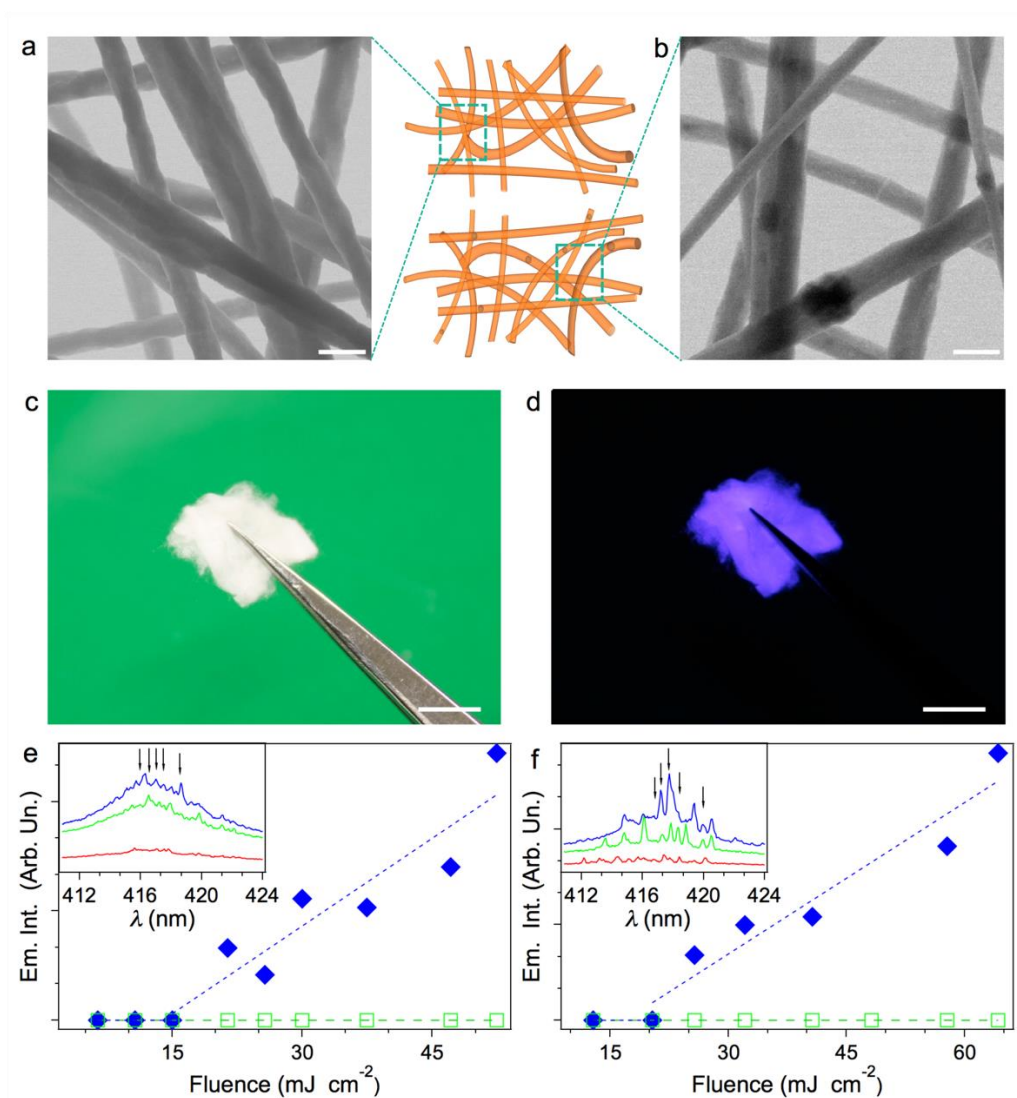

**Figure 1.** Morphology and emission characteristics of Fl-Cz-Fl/PS fiber random lasers. (a, b) Scanning transmission electron micrographs of Fl-Cz-Fl/PS fibers without (a) or with (b) $TiO_2$ particles. Scale bar: 5 μm. Schemes of the obtained structures are also shown. (c, d) Mats of Fl-Cz-Fl/PS fibers under white (c) and UV illumination (d). Scale bar: 1 cm. (e, f) Laser emission intensity vs. excitation fluence for devices without (e) or with (f) $TiO_2$ (full diamonds). Empty squares: PL intensity. Insets: emission spectra for increasing fluences (from bottom to top, e: 26.7, 46.0, and 53.5 mJ cm$^{-2}$ and f: 24.6, 44.9, and 64.2 mJ cm$^{-2}$). Vertical arrows in the insets point peaks at the wavelengths considered for correlation analysis.





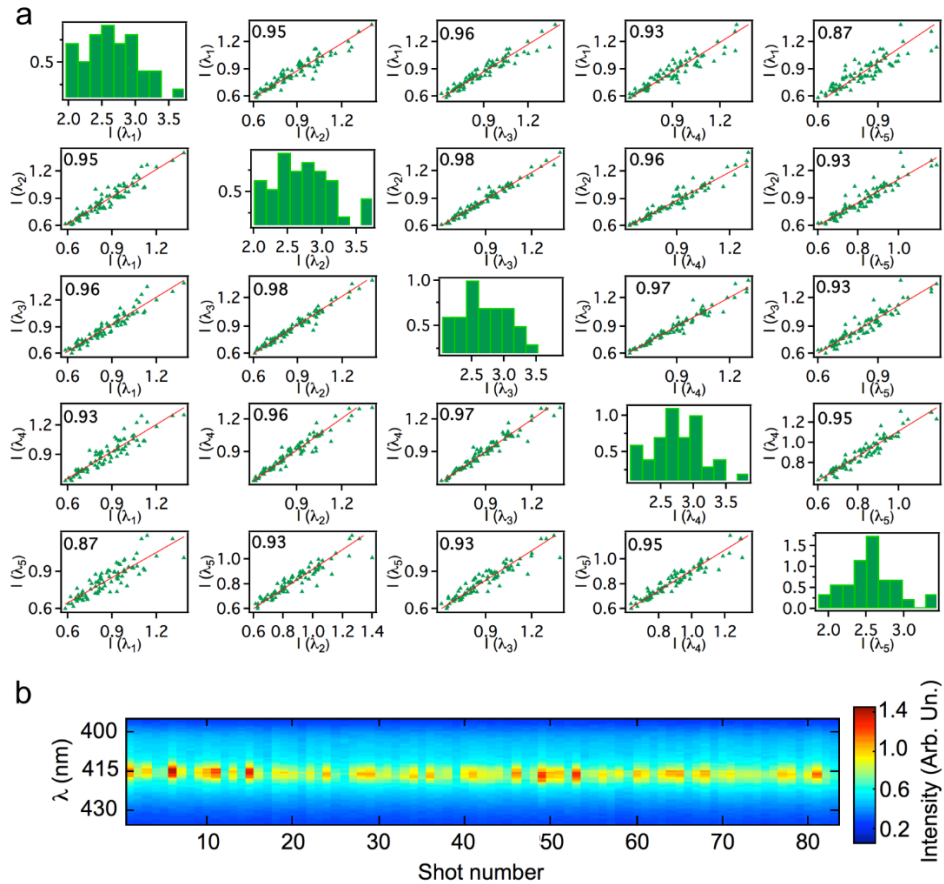

**Figure 2.** (a) Matrix of plots showing the correlation among pairs of emission intensities, $I(\lambda_j)$ and $I(\lambda_k)$, for each pair of lasing wavelengths ($\lambda_j$, $\lambda_k$) shown in the insets of Fig. 1e and 1f and measured for Fl-Cz-Fl/PS fibers. $\lambda_1$=415.9 nm, $\lambda_2$=416.6 nm, $\lambda_3$=417.0 nm, $\lambda_4$= 417.5 nm, $\lambda_5$=418.6 nm. Continuous lines are linear fits to the data. Numbers in the top-left angles in each plot are the correlation coefficients. Histograms representing distributions, normalized to probability density, of the intensity measured for each wavelength at varying excitation pulse are displayed along the matrix diagonal. Excitation fluence = 46 mJ cm$^{-2}$. (b) Single-shot emission spectra evolution upon increasing the number of excitation pulses.





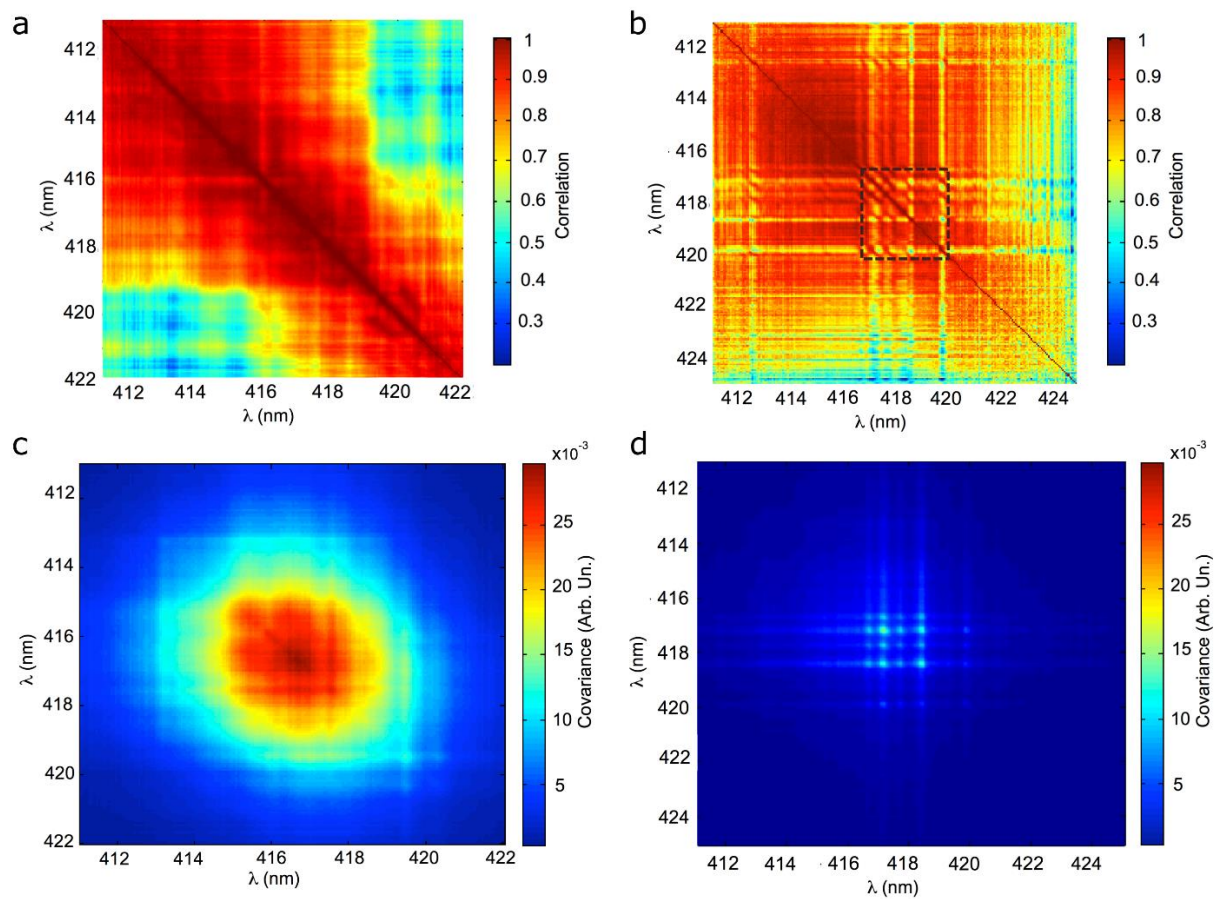

**Figure 3.** (a, b) Map showing the correlation function of $N$ single-shot emission spectra acquired from a sample without (a) or with (b) TiO$_2$ particles, vs. wavelength. $N$=44-83. Excitation fluences = 46.0 mJ cm$^{-2}$ (a), 42.8 mJ cm$^{-2}$ (b). The dashed rectangle in (b) highlights a region with higher internal correlation and lower $C$ values at its vertexes, evidencing reduced inter-mode interactions in TiO$_2$-doped fiber random lasers. (c, d) Corresponding covariance maps.





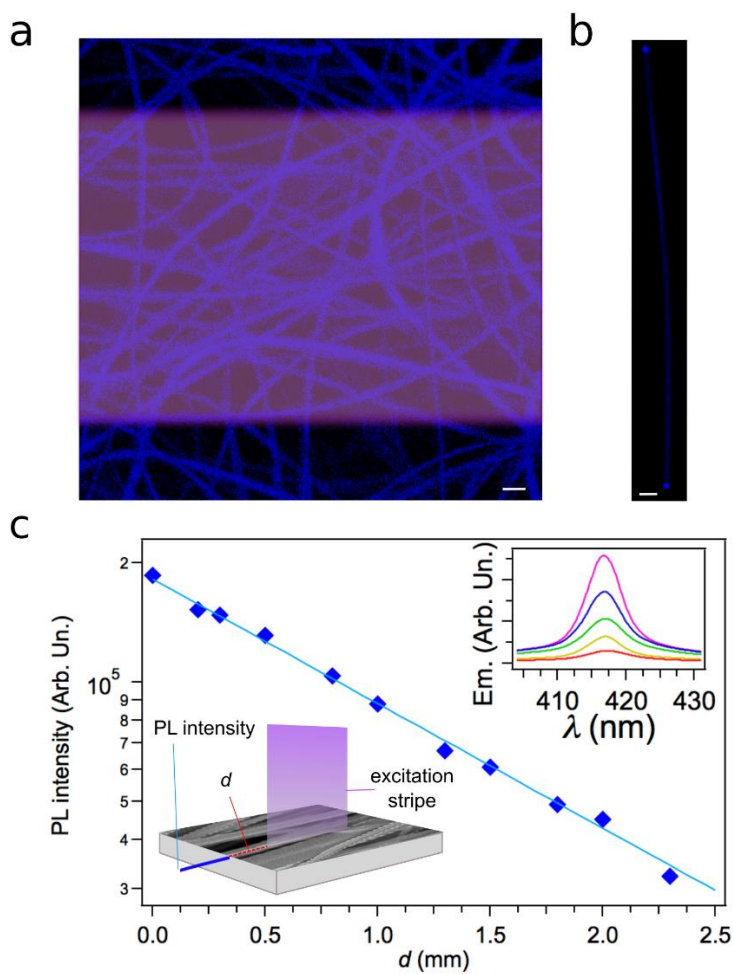

**Figure 4.** Waveguiding properties of Fl-Cz-Fl/PS fibers. (a) Confocal fluorescence micrograph of a mat of random fibers with the superposition of the excitation stripe (violet region, in scale). Scale bar: 5 μm. (b) Fluorescence image of an individual fiber with bright tips highlighting waveguiding of self-emitted light. The sample is excited by a UV lamp. Scale bar: 5 μm. (c) Spatial decay of emitted light intensity (diamonds) guided along Fl-Cz-Fl/PS fibers, deposited on a quartz substrate, as a function of distance, $d$, from photoexcitation. The continuous line is the best fit of the experimental data by an exponential decay (see text). Bottom inset: scheme of the experimental configuration for waveguiding characterization. Top inset: collected spectra for (from top to bottom) $d$=0, 0.3, 0.8, 1.8 and 2.5 mm.





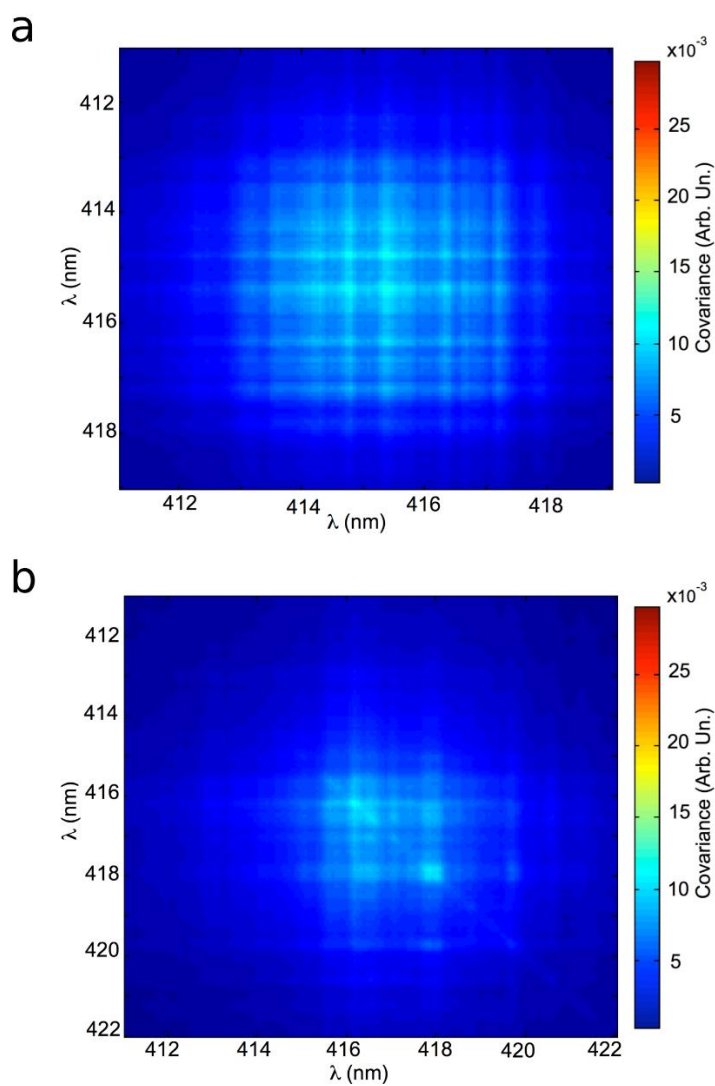

**Figure 5.** Impact of waveguiding on inter-mode intensity correlation. (a) Covariance map for Fl-Cz-Fl/PS fibers embedded in resin. Excitation fluence: 10.7 mJ cm$^{-2}$. Exemplary spectra from the resin-incorporated random lasers are shown in Fig. S5a. (b) Covariance map for Fl-Cz-Fl/PS fibers deposited on quartz, for emission collected with a sample holder-detection optical fiber at an angle of 45° (scheme in Fig. S5b). Excitation fluence: 46.0 mJ cm$^{-2}$.







# Diverse regimes of mode intensity correlation in nanofiber random lasers through nanoparticle doping


Martina Montinaro[1], Vincenzo Resta[1], Andrea Camposeo[2,**], Maria Moffa[2], Giovanni Morello[1],

Luana Persano[2], Karolis Kazlauskas[3], Saulius Jursenas[3], Ausra Tomkeviciene[4],

Juozas V. Grazulevicius[4], Dario Pisignano[2,5,*]

[1] Dipartimento di Matematica e Fisica "Ennio De Giorgi", Università del Salento, via Arnesano, I-73100 Lecce, Italy

[2] NEST, Istituto Nanoscienze-CNR, Piazza San Silvestro 12, I-56127 Pisa, Italy

[3] Institute of Applied Research, Vilnius University, Saulėtekio 3, LT-10257 Vilnius, Lithuania

[4] Department of Polymer Chemistry and Technology, Kaunas University of Technology, Radvilenu Plentas 19, LT-50254 Kaunas, Lithuania

[5] Dipartimento di Fisica, Università di Pisa, Largo B. Pontecorvo 3, I-56127 Pisa, Italy

[*] dario.pisignano@unipi.it

[**] andrea.camposeo@nano.cnr.it


**SUPPORTING INFORMATION**





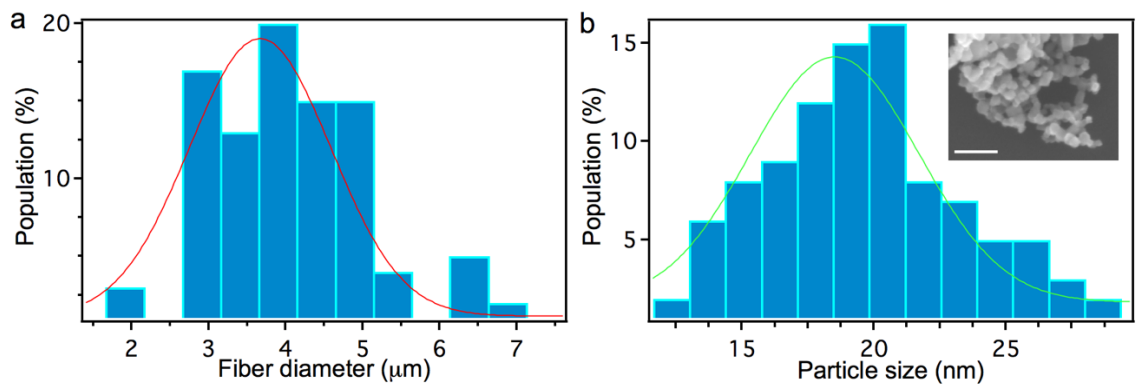

**Figure S1.** (a) Distribution of diameters for Fl-Cz-Fl/PS fibers. The superimposed line is the best fit by a Gaussian curve. (b) Size distribution for TiO$_2$ particles. The superimposed line is the best fit by a Gaussian function. Inset: particles imaged by scanning electron microscopy. Scale bar: 100 nm.





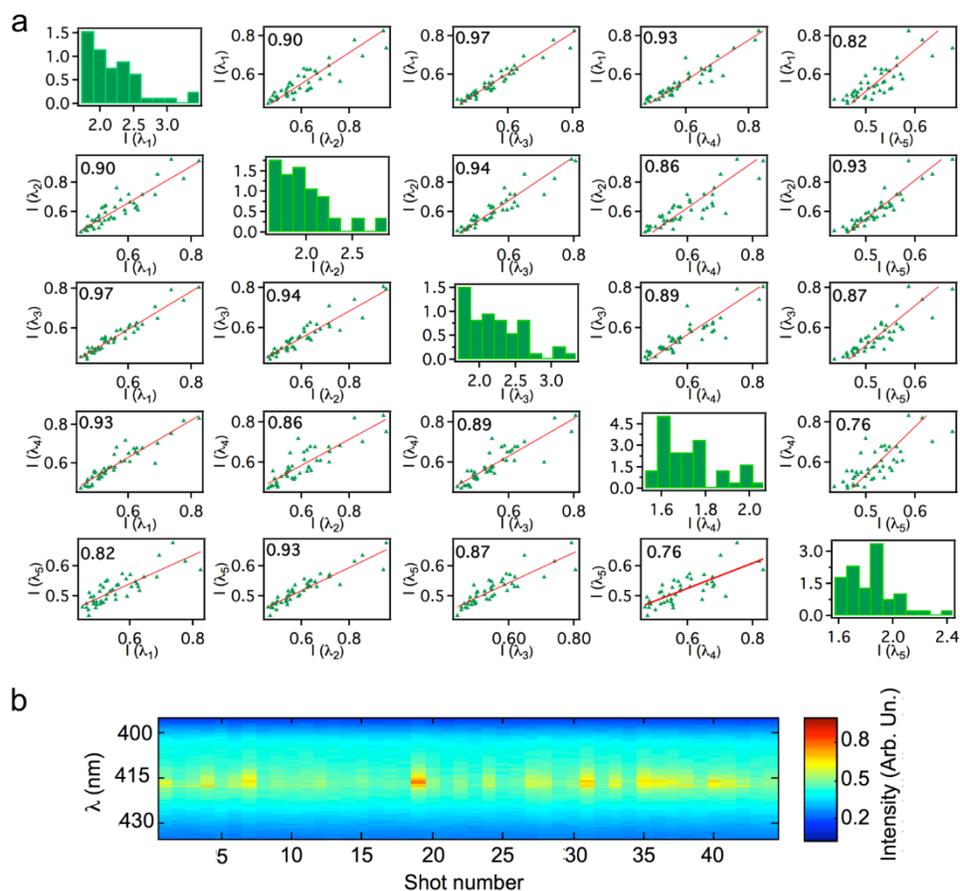

**Figure S2.** (a) Same as Fig. 2, for lasing wavelengths shown in Fig. 1f and measured for Fl-Cz-Fl/PS nanofibers doped with $TiO_2$ nanoparticles. $\lambda_1$=416.7 nm, $\lambda_2$=417.2 nm, $\lambda_3$=417.7 nm, $\lambda_4$= 418.5 nm, $\lambda_5$=419.9 nm. Excitation fluence = 42.8 mJ cm$^{-2}$. (b) Single-shot emission spectra evolution upon increasing the number of excitation pulses.





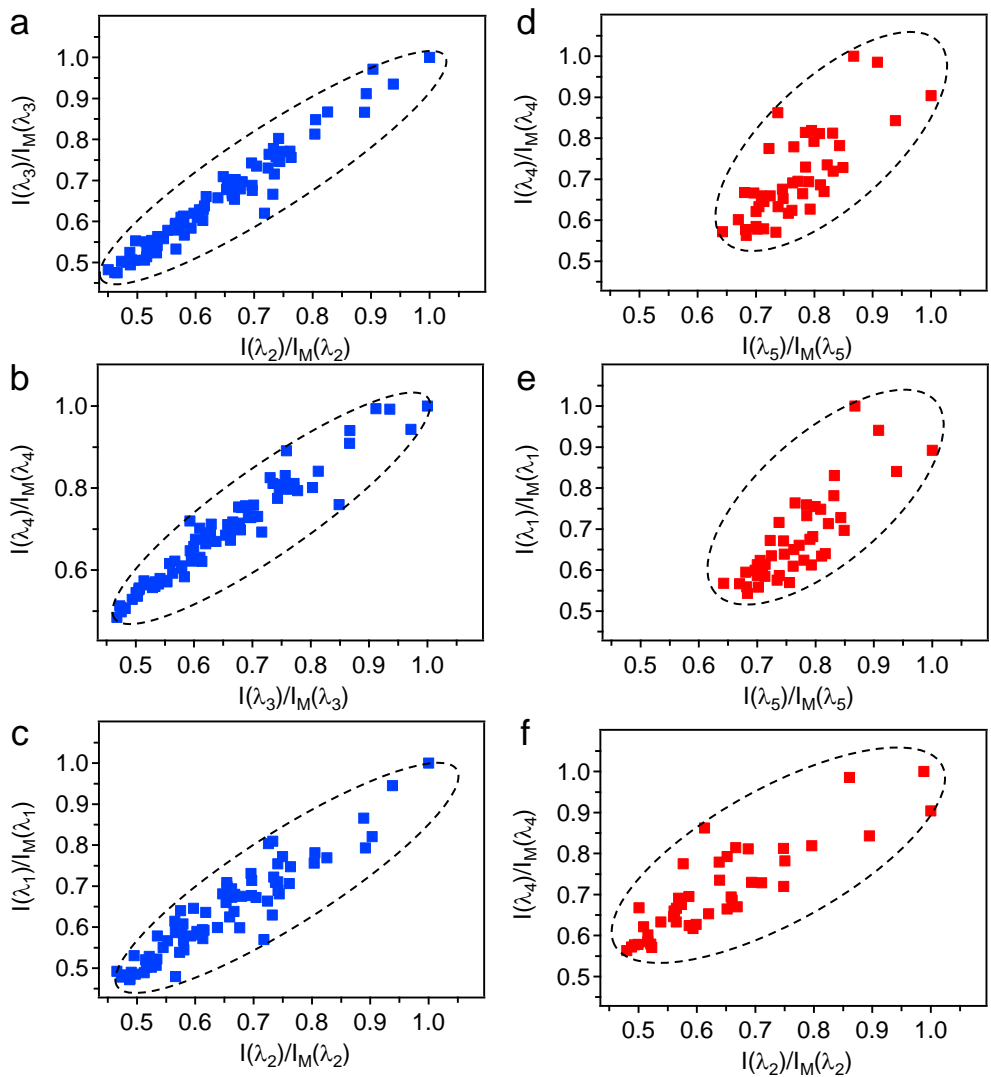

**Figure S3.** Comparison of the correlation of emission intensities for pairs of lasing wavelengths measured for Fl-Cz-Fl/PS fibers (a)-(c) and for Fl-Cz-Fl/PS nanofibers doped with $TiO_2$ nanoparticles (d)-(e). Each set of intensity data $I(\lambda_k)$ is normalized to its maximum value, $I_M(\lambda_k)$. The values of $\lambda_1, \lambda_2, \lambda_3, \lambda_4$ shown in (a)-(c) are those analyzed in Fig. 2, whereas $\lambda_1, \lambda_2, \lambda_4, \lambda_5$ shown in (e)-(f) are those reported in Fig. S2a. The dashed lines are guides for the eyes, highlighting the increased off-diagonal spread of data for samples with $TiO_2$ particles.





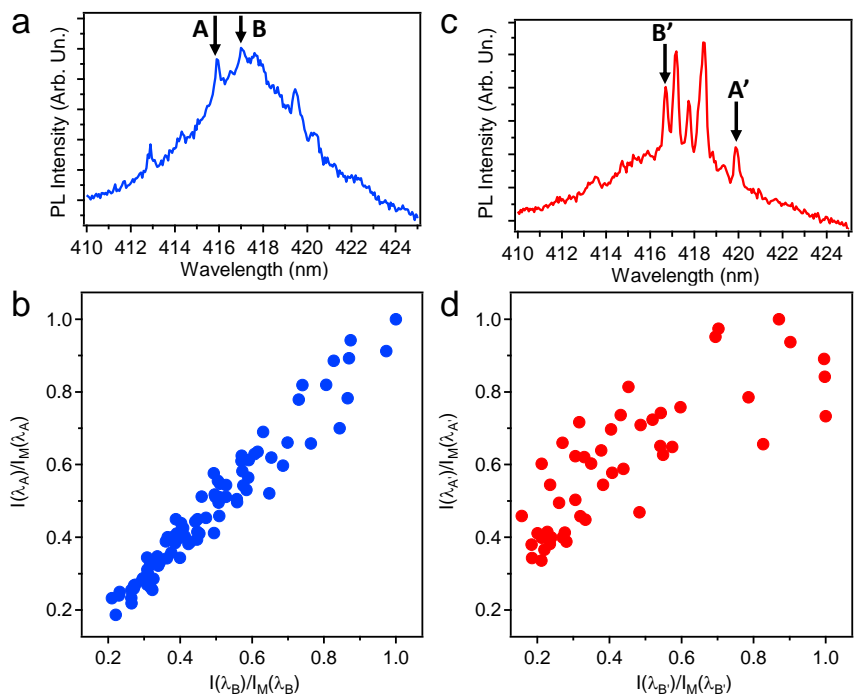

**Figure S4.** Comparison of single-shot emission spectra, (a) and (c), and correlation of emission intensities for pairs of lasing wavelengths, (b) and (d). These are measured for either Fl-Cz-Fl/PS fibers (a, b) or Fl-Cz-Fl/PS fibers doped with $TiO_2$ nanoparticles (c, d). Each set of intensity data $I(\lambda_k)$ shown in (b) and (d) is normalized to its maximum value, $I_M(\lambda_k)$. The spectral position of $\lambda_A, \lambda_B$ and $\lambda_{A'}, \lambda_{B'}$ are highlighted by vertical arrows in (a) and (c), respectively.





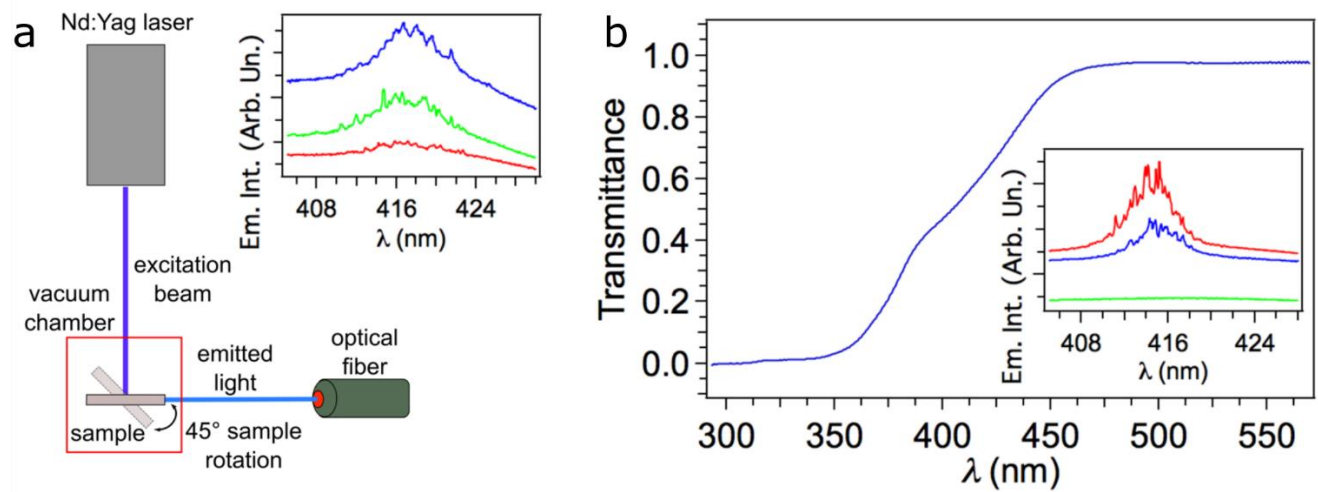

**Figure S5.** (a) Schematics showing sample rotation with respect to the optical fiber collecting emission. Inset: Emission spectra (fluences from bottom to top: 19.3, 46.0 and 61.0 mJ cm$^{-2}$). (b) Transmission spectrum of a reference resin film (NOA 68). At the excitation wavelength (355 nm), transmittance is about 0.05. Inset: Emission spectra of the random lasers embedded in resin (effective excitation fluences, from bottom to top: 5.9, 9.6 and 10.7 mJ cm$^{-2}$).